\documentclass{article}

\usepackage{PRIMEarxiv}

\usepackage[utf8]{inputenc} 
\usepackage[T1]{fontenc}    
\usepackage{hyperref}       
\usepackage{url}            
\usepackage{booktabs}       
\usepackage{amsfonts}       
\usepackage{nicefrac}       
\usepackage{microtype}      
\usepackage{lipsum}
\usepackage{graphicx}
\graphicspath{{media/}}     
\usepackage{multirow} 
  
\title{Combining topic modelling and citation network analysis to study case law from the European Court on Human Rights on the right to respect for private and family life
}

\author{M. Mohammadi  \\
        University of Groningen\\
        The Netherlands\\
        \texttt{mohammadimathstar@gmail.com}\\
        \And
        L. M. Bruijn \\
        University of Groningen\\
        The Netherlands\\
        \texttt{l.m.bruijn@rug.nl}\\
        \And
        M. Wieling \\
        University of Groningen\\
        The Netherlands\\
        \texttt{m.b.wieling@rug.nl}\\
        \And
        M. Vols\\
        University of Groningen\\
        The Netherlands\\
        \texttt{m.vols@rug.nl}\\
}

\begin{document}
\maketitle

\begin{abstract}
As legal case law databases such as HUDOC continue to grow rapidly, it has become essential for legal researchers to find efficient methods to handle such large-scale data sets. Such case law databases usually consist of the textual content of cases together with the citations between them. This paper focuses on case law from the European Court of Human Rights on Article 8 of the European Convention of Human Rights, the right to respect private and family life, home and correspondence. In this study, we demonstrate and compare the potential of topic modelling and citation network to find and organize case law on Article 8 based on their general themes and citation patterns, respectively. Additionally, we explore whether combining these two techniques leads to better results compared to the application of only one of the methods. We evaluate the effectiveness of the combined method on a unique manually collected and annotated dataset of Aricle 8 case law on evictions. The results of our experiments show that our combined (text and citation-based) approach provides the best results in finding and grouping case law, providing scholars with an effective way to extract and analyse relevant cases on a specific issue.
\end{abstract}


\section{Introduction}
\label{intro}
Case law plays a crucial role in legal research, particularly in the context of human rights. Many international human rights conventions, such as the European Convention on Human Rights (ECHR), are considered `living instruments', which means that human rights should be interpreted in light of present-day conditions and in accordance with developments in international law \cite{letsas2013echr}. Fundamental human rights, such as the right to respect for private and family life, home, and correspondence as enshrined in Article 8 of the ECHR, serve as broad normative standards that (may) evolve in response to societal changes and international consensus. For example, the meaning of `correspondence' has significantly changed with the internet and the progression of technology, and also what is considered `family life' \cite{wray2023article} or a `home' is ever-developing \cite{bec610359de944ea828cfbe71a60146d}. Consequently, the interpretation and application of human rights undergo continuous development, requiring legal scholars and practitioners to rely heavily on the case law established by international courts, such as the European Court of Human Rights (ECtHR). 

However, the volume of case law is ever-increasing, which makes it challenging for legal scholars to discover relevant cases and gain a comprehensive understanding of this vast amount of information. To overcome this challenge, the application of computational and automated methods may offer valuable solutions. In other domains, such as social sciences, physics, and computer science, data-driven approaches such as Machine Learning and Natural Language Processing have proven highly effective in analyzing and extracting meaningful insights from large body of texts \cite{trompper2016automatic,livermore2017supreme,shulayeva2017recognizing,lippi2019claudette,ruggeri2022detecting,schepers2023predicting,medvedeva2020using,medvedeva2021automatic,medvedeva2022rethinking}.

Two techniques that hold particular promise to make sense of all the available case law are topic modelling and citation network analysis. Topic modelling is a computational technique that allows for the identification and clustering of multiple topics (i.e.~subjects) in a collection of documents \cite{remmits2017finding,dyevre2021text,silveira2021topic,razon2022topic,luz2020topic}. Topic modelling defines a topic as a group of related words that frequently appear together within a given text. For example, in case law about housing issues, we would anticipate the frequent appearance of terms such as `apartment', `landlord', `eviction', and `rental agreement'. These words can be grouped together into a cohesive topic (e.g., `housing'), shedding light on the underlying theme or subject matter. Interestingly, despite being a well-established technique in computer science, topic modelling does play a relatively minor role in legal research (see Section \ref{sec:related_work}). 

Citation network analysis in a legal context can be used to analyse the interplay of citations between legal documents, such as court decisions. By tracing the web of references among court decisions, network analysis is able to unveil distinct patterns within citation networks. For example, cases sharing common themes may tend to exhibit a higher volume of interconnections through citations when compared to cases with unrelated topics. Consequently, by examining citations between court decisions, researchers could potentially identify the content of cases, the relationships between cases, evolving patterns in case law, and identify the landmark cases that hold a central position in the network \cite{fowler2008authority,knops2014holding,christensen2016identification,whalen2016legal,frankenreiter2017network,olsen2017finding,derlen2017good,vsadl2017can,arnold2023scaling,gorski2021network,renberg2021mapping}. 

Although citation network analysis has received more attention in legal research when compared to topic modelling (see Section \ref{sec:related_work}), the usefulness of two approaches to organizing and studying case law has not been compared. In addition, it has not been analysed whether combining the textual content of cases identified through topic modelling with their interconnectivity identified via citation analysis could yield a more comprehensive understanding of the case law.

In this study, we aim to fill this gap by demonstrating and comparing the potential of topic modelling and citation network analysis to collect and organize a large collection of case law. Additionally, we explore whether combining these two techniques leads to an enhanced understanding of the data compared to the application of only one of the methods. We evaluate our approach in the context of the case law of the ECtHR concerning the right to private and family life, home, and correspondence (Article 8 of the ECHR). This is a broad and multifaceted right that covers a wide range of issues, from data protection to housing, and from family reunification to adoption \cite{brems2015don}.  Hence, the body of case law spans many different contexts, which makes it challenging to identify overarching topics, trends or patterns via conventional doctrinal legal research methods \cite{vols2021legal}. 

The objective of this paper is therefore to develop and test an enhanced computational method to collect and organize case law. 
To achieve this, we conduct a series of three experiments. The first experiment applies topic modelling (using Latent Dirichlet Allocation or LDA) to uncover (hidden) topics in all ECtHR case law written in the English language on Article 8 (\textit{n}=6,854) and to cluster the cases with similar topics. 
Having access to a list of eviction-related judgements and decisions, we then use LDA's output to determine how cases on the specific issue ''eviction'', which refers to the involuntary loss of one's home \cite{vols2023optional,sweeney2023deconstructing}, fit into the broader landscape of Article 8 case law. 

The second experiment applies citation network analysis to detect communities (using the Louvain algorithm) in the network of case law on Article 8. Communities are formed based on the citation patterns among cases. Given the vast number of cases, it is not feasible to conduct a thorough analysis of each case and its community. Therefore, our attention is directed towards the communities that involve eviction cases.

In the third experiment, we will combine both methods. In other words, we integrate the textual content of the cases with their interconnections based on citations. We investigate the potential of this integration for case retrieval, meaning the successful detection of cases on a specific issue, eviction.

In the following section, we discuss previous work concerning the application of topic modelling and network analysis in the legal domain. In Section \ref{sec:data}, we describe the data we have used for our experiments and how we collected the data, both the data set with Article 8 case law, the manually collected data set of eviction cases, and the data set with the citation network of Article 8. In Sections \ref{sec:topic_modelling}, \ref{sec:net_analysis}, and  \ref{sec:case_retrieval}, 
we describe our three experiments and report the results. In Section \ref{sec:conclusion}, we conclude our work by summarizing the key findings and discussing their implications for future research and practical applications.

\section{Related work}\label{sec:related_work}

Machine Learning (ML) and Natural Language Processing (NLP) have emerged as transformative technologies with the potential to revolutionize various fields, including legal research. ML involves the development of algorithms that enable computers to learn from data and make predictions or take actions without being explicitly programmed. NLP focuses on the interaction between computers and human language, allowing machines to understand, interpret, and generate human language in a way that resembles human communication. 

ML and NLP have already made significant contributions to the field of legal research. They have been used to classify relevant entities in court documents \cite{dozier2010named,cardellino2017low,schroeder_lindholm_2023}, to automatically summarise case law \cite{pandya2019automatic,kanapala2019text,galgani2015summarization,kumar2012legal,kim2013summarization} and to identify, categorise and forecast court decisions \cite{aletras2016predicting,medvedeva2020using,medvedeva2021automatic,medvedeva2021automatically,medvedeva2023rethinking}. These approaches not only address practical challenges in the legal discipline but also broaden and develop legal research methodology.

A powerful tool in the NLP toolbox for analyzing legal data is topic modelling. While relatively new in legal research, topic modelling is a popular computational technique used to extract (latent) themes or topics from large collections of textual data. It is applied to various tasks such as document clustering, text classification, sentiment analysis, and information retrieval \cite{asmussen2019smart,li2021bibliometric}. 
In the realm of legal research, topic modelling has been applied to analyze case law from different jurisdictions, including the Australian High Court \cite{carter2016reading}, the Supreme Court of the Netherlands \cite{remmits2017finding}, the Brazilian Supreme Court \cite{luz2020topic,aguiar2022using}, the Czech Supreme Court \cite{novotna2020topic}, the United States Supreme Court \cite{silveira2021topic}, the Philippine Supreme Court \cite{razon2022topic}, and the Housing law tribunal in Canada \cite{salaun2022tenants}.  Other studies in the legal domain have applied topic modelling to classify individual judicial opinions at the International Criminal Court \cite{wigard2023matter}, legislative texts from the United Kingdom \cite{o2016analysis} and Latvian legal acts \cite{viksna2020exploring}. 

Another computational method for dealing with legal data is citation network analysis, which investigates the citation patterns within legal documents. In contrast to topic modelling, citation network analysis is relatively more developed and widely used in the legal domain. It is frequently employed to study the network of citations between court cases e.g.,  \cite{fowler2008authority,knops2014holding,christensen2016identification,whalen2016legal,frankenreiter2017network,olsen2017finding,derlen2017good,vsadl2017can,arnold2023scaling,gorski2021network,renberg2021mapping}. Both national and international courts often cite previous court decisions to support their own decision or to distinguish or overrule previous cases. These citations between cases can create a network by connecting cases; when a case cites another case, a link is created between them. Over time, as more cases cite other cases, the network becomes more complex and interconnected \cite{vsadl2017can}. By applying citation network analysis, researchers gain insights into the structure and dynamics of (part of) a legal system. Citation network analysis could, for example, be applied to study if, how, and why courts reference other decisions, to identify important court decisions (i.e.~highly cited decisions), and to track the evolution of legal doctrines over time \cite{olsen2017finding}.

Some studies used network science to analyse or predict future citation behaviour. For instance, Leitão and colleagues analyzed case law from the ECtHR to study how cases reach their current number of citations, model the evolution of citations and predict future citations \cite{leitao2019quantifying}. Similarly, Mones and colleagues focused on predicting new citations within the network of the case law of the CJEU \cite{mones2021emergence}. Schepers and colleagues conducted a similar study using case law from the Netherlands \cite{schepers2023predicting}. Their research aimed not only to predict the citation of court decisions by other courts but also to uncover the key factors that determine whether a case would be cited or not. These studies highlight the potential of network science to provide insights into the dynamics of legal citation networks and anticipate future citation behaviour.

With this paper, we aim to contribute to the existing body of literature by comparing and combining topic modelling and citation network analysis to organize and analyze a large collection of case law. 

\section{Data}\label{sec:data}

\subsection{Case law on Article 8 of the ECHR}
For the purpose of this paper, we collected all ECtHR case law (judgments and decisions) on Article 8 of the ECHR. This provision protects the right to respect for private and family life, covering a wide range of issues from family reunification to adoption, and from data protection to housing. 

All case law of the ECtHR is available in the HUDOC (HUman rights DOCument) database.\footnote{\href{https://www.echr.coe.int/hudoc-database}{https://www.echr.coe.int/hudoc-database}} HUDOC provides free online access to a vast collection of legal materials related to the ECtHR's work. It includes judgments, decisions, advisory opinions, reports, and other documents issued by the court. The database dates back to the establishment of the ECtHR in 1953. Through HUDOC, users can search for specific cases, access full-text documents, and retrieve information about the Court's judgments and decisions. The database allows users to search based on various criteria, such as keywords, parties involved, legal provisions, and the date of the decision. 

We collected all case law on Article 8 until January 2023,\footnote{The code for downloading data can be found in this \href{https://github.com/WillSkywalker/HUDOCcrawler}{repository}.} including the metadata of these cases, which consists of the name of the case, the originating body, the type of document (judgment or decision), the application number, citations, and the importance level (as provided by HUDOC). This resulted in a total of 9,777 cases. Of these cases, 6,854 are in English and 2,923 in French, the two official languages of the ECtHR. The distribution of Article 8 case law, differentiating between English and French cases, as well as judgments and decisions, is presented in Table \ref{tab:article8-summary}.


The ECtHR issues two types of rulings: decisions and judgments. Decisions predominantly pertain to the admissibility of a case, determining whether it meets the necessary procedural requirements to proceed further. If a case is deemed admissible, it advances to the judgment phase. In judgments, the ECtHR evaluates the merits and ultimately decides whether a human right has been violated. Decisions have historically received less attention in legal research. However, their significance has grown notably in recent years, driven by the ECtHR's response to mounting caseloads, the imperative for more efficient case management and political backlash \cite{graham2020strategic,glas2023age}. The ECtHR has put more emphasis on the admissibility criteria and uses decisions to clarify the boundaries and interpretations of human rights, offering valuable insights into the Court's evolving approach to human rights cases. For instance, decisions such as \textit{FJM v.~the United Kingdom} on the right to housing and \textit{Le Mailloux v.~France} regarding the right to life have offered significant contributions to the interpretation and application of these rights even though the ECtHR deemed their application inadmissible \cite{fick2022horizontality}.\footnote{ECHR 6 November 2018, no. 76202/16 (\textit{FJM v.~the United Kingdom}); ECHR 5 November 2020, no. 18108/20 (\textit{Le Mailloux v.~France})}

Therefore, our study includes both judgements and decisions to ensure a comprehensive analysis of Article 8 case law. This not only makes our research innovative because of the (combination of) methods, but also because of the inclusion of the decisions, which have been often overlooked in prior studies. 
\begin{table}[h]
    \centering
    \begin{tabular}{c c|c c}
        \hline
         \multicolumn{2}{c|}{English} & \multicolumn{2}{|c}{French} \\
         \hline
         Decisions & Judgments & Decisions & Judgments \\
         4,839 & 2,015 & 2,274 & 649\\
         \hline
    \end{tabular}
    \caption{Total number of cases about Article 8 ECHR}
    \label{tab:article8-summary}
\end{table}

\subsection{Case law on eviction}
To assess whether our topic modelling and citation network analysis are able to identify relevant and related clusters of case law, we specifically focused on a subset of cases involving evictions. For this purpose, a team of human coders aimed to manually identify all Article 8 decisions and judgements in which the applicant filed a complaint concerning his/her eviction. We use this dataset to evaluate whether all manually identified eviction cases belong to the same computer-identified cluster in the network of Article 8 case law.

To manually identify all Article 8 decisions and judgments relating to eviction, we searched the HUDOC database filtering on Article 8 cases, and we used the keywords search option to search for the following terms: housing, home, house, eviction, homelessness, evictee, tenant, tenants, rent, ``right to adequate housing'', landlord, landlords, and Article 8. After the keyword search, we manually identified the relevant cases, which are both judgments and decisions about Article 8 of the ECHR in which the applicant complains about his/her eviction. We employ a rather broad interpretation of the term eviction (i.e.~the involuntary loss of one's home), including the destruction of someone's house (i.e. demolition). This search resulted in a total of 198 eviction cases, consisting of 104 decisions and 94 judgments until January 2023. 

\subsection{Citation network}\label{subsec:cite_net}
As we would like to analyze the citation behaviour within the network of Article 8 case law, we need to find citations among the cases. To achieve this, we leveraged the available meta-data in HUDOC to extract citations between cases. Subsequently, we utilized \href{https://networkx.org/}{NetworkX} (a network analysis library) to construct a citation network encompassing all references among cases on Article 8. The resulting network comprises 9,777 nodes representing individual judgements or decisions (both English and French), interconnected by 39,582 links. 
This network provides a comprehensive overview of the cases about Article 8 and their interrelationships.

The resulting network consists of 2,464 subnetworks (i.e.~separate smaller networks with no links between any pair of them). Consequently, most subnetworks only contain only one or just a few nodes. A total of 2,408 subnetworks contain just a single node (i.e. a singular case). And 55 subnetworks contain up to 8 nodes\footnote{45 subnetworks contain 2 nodes; four subnetworks contain 3 nodes; three subnetworks contain 4 nodes; two subnetworks contain 6 nodes; one subnetwork contains 8 nodes.}. However, there is one single, very large subnetwork (henceforth dubbed the ''giant subnetwork''), encompassing 7,234 nodes (constituting 74\% of the entire network). It's noteworthy that this large subnetwork also comprises 86.8\% of the eviction cases (171 out of 198 cases). Given that all other subnetworks are comprised of only one node and a few up to eight nodes, we only focus on this large subnetwork in Experiments 2 and 3.


\section{Experiment 1: topic discovery and case clustering }\label{sec:topic_modelling}


In Experiment 1, we conduct topic modelling to identify the hidden topics in our data set and to cluster the cases based on their shared subject matter. We first conduct the experiment on all case law in English pertaining to Article 8 of the ECHR (\textit{n}=6,854). Subsequently, we assess the specific topics associated with our manually collected subset of eviction cases (\textit{n}=198). We hypothesize that this subset spans a spectrum of diverse eviction types, such as evictions due to urban planning, non-payment of rent or due to criminal activities, thus potentially spanning across multiple topics.

\subsection{Method}
To extract topics within the case law of Article 8, several techniques have been proposed, such as Non-negative Matrix Factorization (NMF; \cite{obadimu2019identifying}), Latent Dirichlet Allocation (LDA; \cite{blei2003latent}), Top2Vec \cite{angelov2020top2vec}, and BERTopic \cite{grootendorst2022bertopic}. As these methods are so-called unsupervised techniques\footnote{In unsupervised techniques, the model operates without human supervision, which means that no data is provided which has the desired outcome or labels.}, they are suitable for exploring major trends within large corpora of texts.
For this study, we selected Latent Dirichlet Allocation (LDA). This technique is recognized for its transparency and the incorporation of uncertainty modelling, which facilitates the interpretation of the extracted topics. Moreover, in contrast to methods such as BERTopic and Top2Vec, LDA allows a document to contain multiple topics, which is particularly advantageous in capturing the complexity and diverse nature of case law.\footnote{The judicial decisions in the European Court of Human Rights (ECtHR) are comprised of multiple sections that address different aspects of the case. The main sections include a) Procedure: This section presents the case's procedural history. b) Facts: This section provides relevant background information that led to the case being presented to the court and any relevant laws (excluding ECtHR). c) Law: This section presents the legal argument of the court. d) Judgment: This section presents the court's final decision regarding the alleged violation.} 
These features have increased LDA's popularity as a powerful topic modelling tool in various domains, including  NLP and information retrieval. While the exact functioning of LDA is quite involved (see \cite{blei2003latent} for all details), the main characteristic is that it is a probabilistic method in which the distribution of words (in the individual documents) is used to group words into topics. A single word may be associated with different topics (with different probabilities). One or more topics will be associated to each document on the basis of the words occurring in the document. It is not necessary to provide a list of topics beforehand. Instead, the algorithm will automatically group a set of related words into a topic. Of course, interpreting the constructed topics does require human involvement.  


\subsection{Results: topic discovery and clustering of cases on Article 8 of the ECHR}
In Experiment 1, we apply LDA to the collection of all 6,854 English decisions and judgments of the ECtHR on Article 8. We first normalize the texts to prepare the court decisions for our analysis using spaCy \cite{spacy2}. The normalization process involves the following five steps. Step one is tokenization: segmenting the texts into smaller units called tokens, which can be individual words, punctuation, and other elements such as numbers. Step two is to remove any punctuation, common words,\footnote{Words that are frequently used in texts but have little value for analyzing texts, such as `the', `and', `a', `of', and `is'.} numbers, and dates to eliminate noise. Step three is to delete people's names and locations (including names of countries) to avoid bias. Step four is lemmatization: transforming words to their base form. For instance, `run', `running', `ran', and `runs' become `run' (the lemmatized word). Step five is to convert all characters to lowercase for consistency. This whole process transforms the unstructured text of a case into a list of words (i.e.~tokens).

Finally, we use the bag-of-words model to transform lists of tokens into a format suitable for LDA. This bag-of-words model results in vector representations, where each position in the vector corresponds to a unique word, and the value in that position represents the frequency of the word in the document. Each individual word therefore has a single non-zero value. Once the preprocessing step is complete, we use the LDA implementation available in the software library Gensim \cite{rehurek_lrec}. 

In LDA, the user must specify the number of topics in the corpus beforehand. Finding the optimal number can be challenging, as it usually entail to vary the specific number and select the one yielding the most coherent and interpretable topics. We adopt a mixed-methods approach combining quantitative and qualitative evaluations of the identified topics to address this issue. We first use coherence scores (i.e.~how similar the top words associated with a topic are; see \cite{syed2017full} for more details) to quantitatively assess the coherence of each LDA model. Then, we manually check the cohesion of the extracted topics for various candidate models (with higher coherence scores). Through this procedure, we select the most interpretable and coherent model. 

\begin{figure}
    \centering
    \includegraphics[scale=0.4]{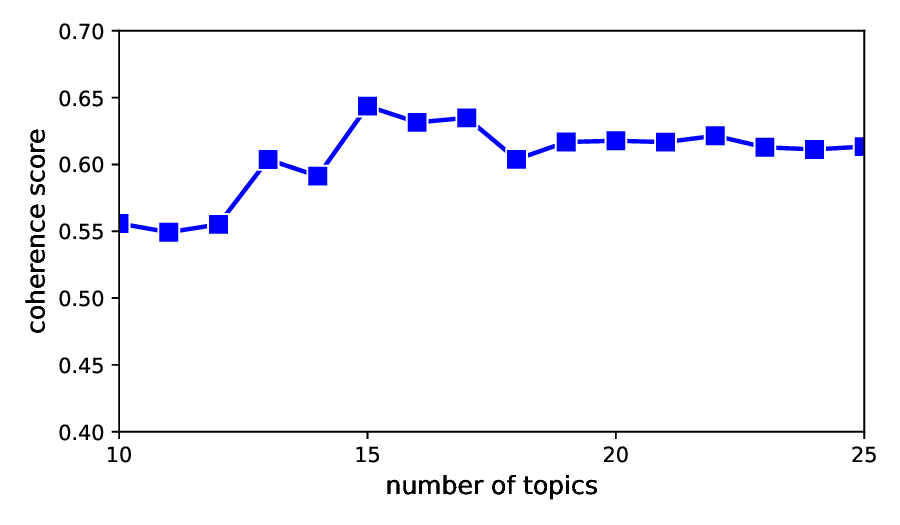}
    \caption{Coherence scores capturing semantic similarity between the most prominent words for each topic.}
    \label{fig:coherence_scores}
\end{figure}

Following the above procedure, we first computed the coherence scores for different numbers of topics, ranging from 10 to 25. Figure \ref{fig:coherence_scores} illustrates the performance of LDA models in terms of coherence scores topics for an increasing number of topics. It shows that the range of 15 to 18 topics yields the highest coherence scores. We then used human judgment to examine the identified topics within this range. Specifically, we inspected the prominent words\footnote{Words with a higher likelihood of occurrence within a specific topic compared to others.} within each topic to evaluate the cohesiveness of the topics. We identified that a set of 17 topics resulted in the most informative topics with minimal overlap between them.  

\begin{figure}[h]
    \centering
    \includegraphics[scale=0.38]{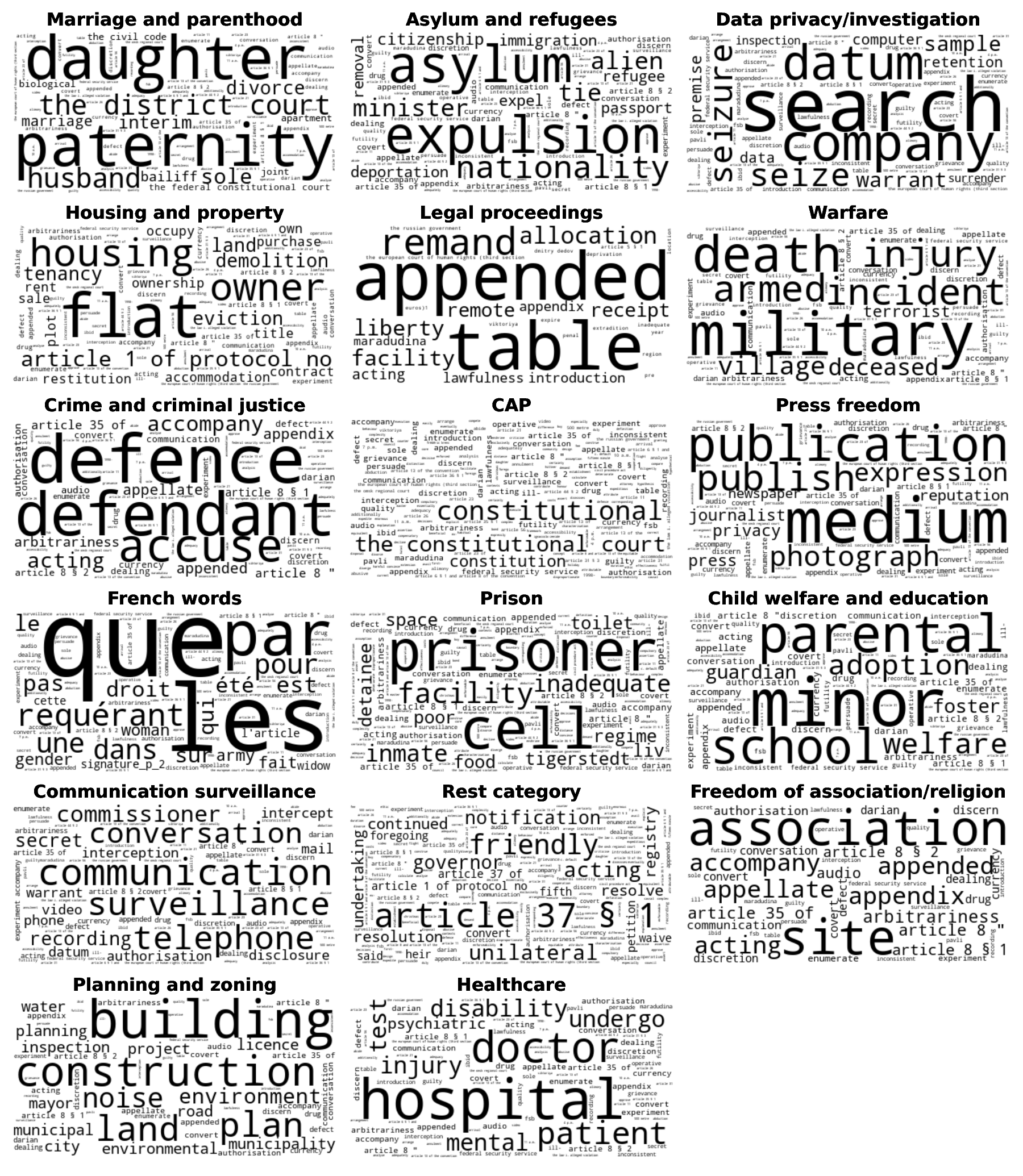}
    \caption{Overview of the most prominent words within each topic where the size of a word captures the likelihood of its occurrence within a topic. Note that CAP is an abbreviation for Constitutional and administrative proceedings.}
    \label{fig:topwords_all}
\end{figure}

Figure \ref{fig:topwords_all} reveals the most prominent words associated with these 17 topics. Based on these words, we manually assigned descriptive labels to each topic (headers of the individual word clouds). The varying set of labels provides an immediate overview of the broad and multifaceted nature of the case law on Article 8 of the ECHR.\footnote{Note that some English decisions and judgements also contain French words. Specifically, for older cases, the documents may contain a French translation. This leads to some French words being included in a topic.}


\begin{figure}[h!]
    \centering
    \includegraphics[scale=0.42]{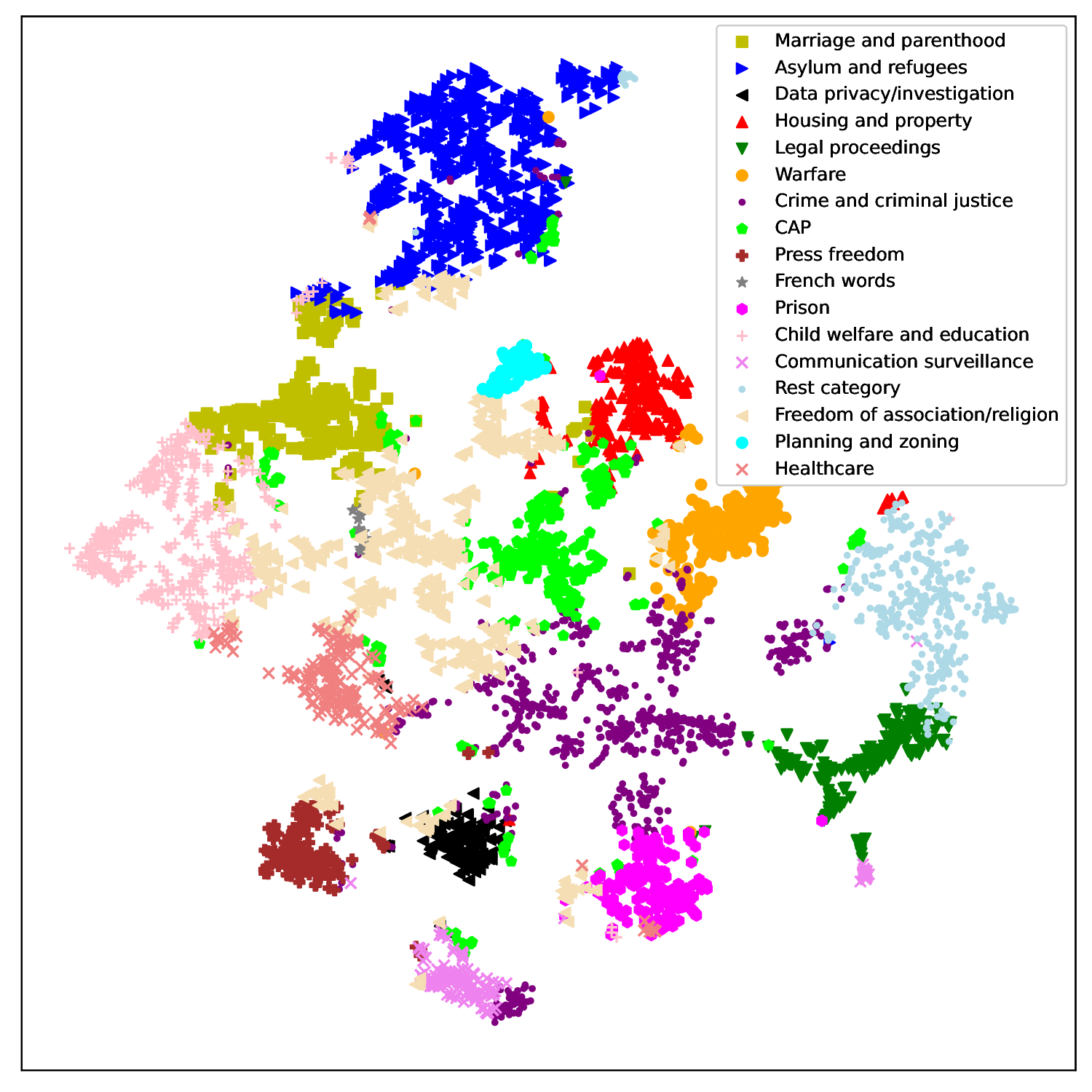}
    \caption{Visualization (via t-SNE) of case laws based on their topics. A color is assigned to each case indicating the most significant topic within the case.}
    \label{fig:tsne-with-topic-color}
\end{figure}


In order to evaluate the effectiveness of the models in grouping cases, we utilize t-distributed Stochastic Neighbor Embedding (t-SNE; \cite{van2008visualizing}), a well-known visualization technique. This technique aims to preserve the similarity between data points while projecting high-dimensional data into a two-dimensional space. In particular, it focuses on keeping cases with similar topics close to each other. This helps to detect clusters in the data that may be difficult to see in the original 17D topic space. 

Figure \ref{fig:tsne-with-topic-color}  presents the visualization generated by t-SNE. Each case is represented in this visualization by a distinct colour and symbol, reflecting its primary topic. This figure shows that cases with similar primary topics mostly tend to form cohesive clusters that can be observed by the concentration of colours and symbols. 
These results illustrate LDA's potential in organizing case law based on the specific topics they address, offering a powerful tool for organizing legal data and enhancing our understanding of legal content and its thematic associations.

\subsection{Results: topic discovery and clustering of eviction cases}
Next, we use the outcomes of Experiment 1 to explore the position of the manually collected eviction cases within the broader landscape of Article 8 case law. Additionally, we aim to identify the prevalent topics in the known eviction cases. 
This will allow us to categorize all eviction-related cases within our annotated data set.

\begin{figure}
    \centering
    \includegraphics[scale=0.35]{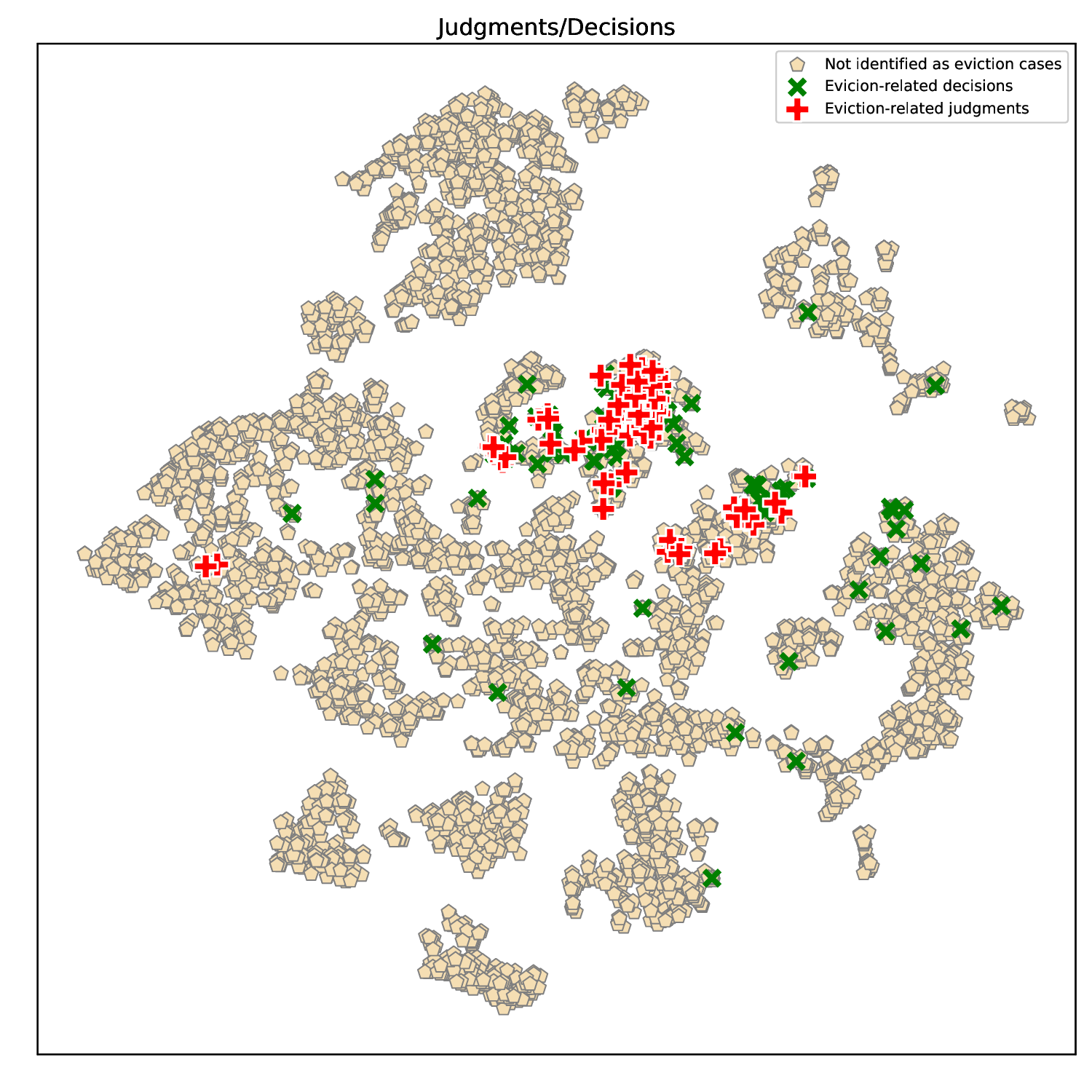}
    \caption{Visualization of LDA's output, using t-SNE, for both judgments and decisions. 
    }
    \label{fig:tsne}
\end{figure}

\begin{figure}
    \centering
    \includegraphics[scale=0.43]{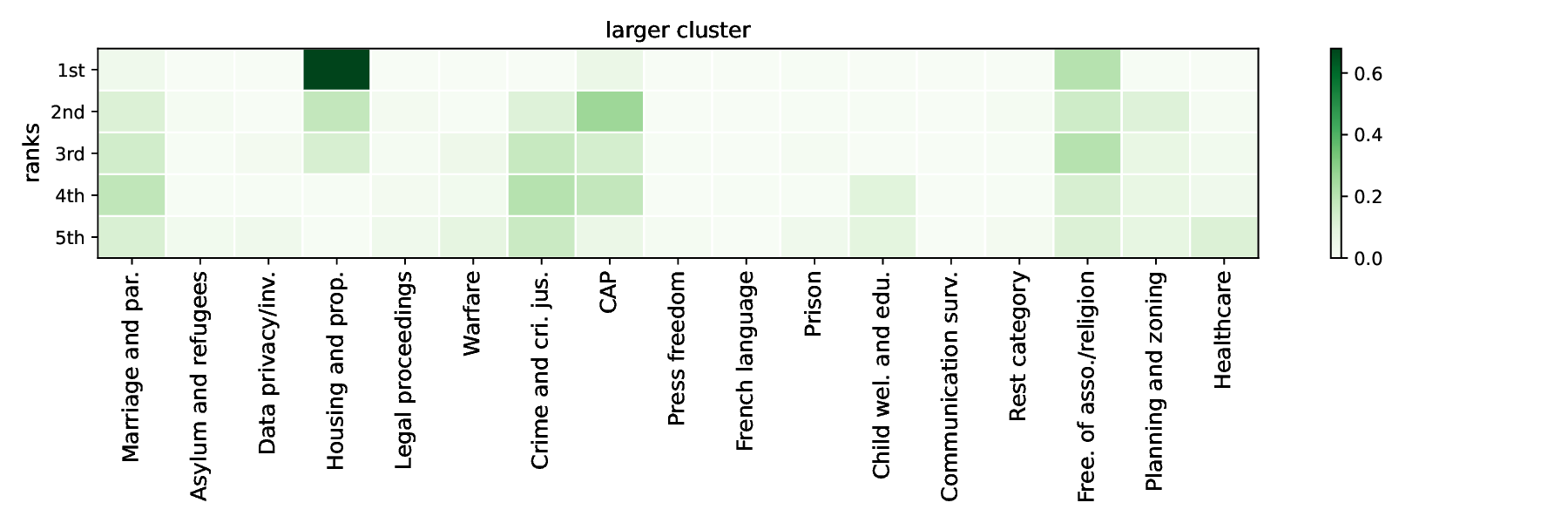} 

    \vspace{-0.5cm}
    \includegraphics[scale=0.43]{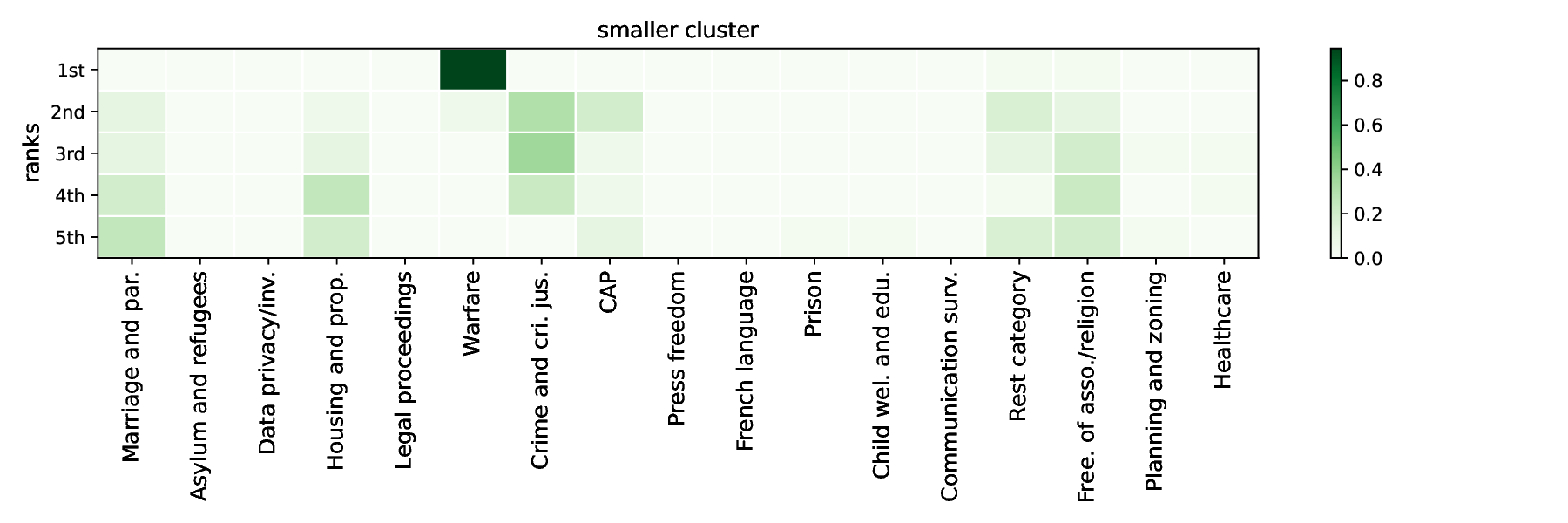}
         
    \caption{
    Topic distribution within eviction-related clusters.
    }
    \label{fig:topic_hist_evict_clusters_all}
\end{figure}

Utilizing t-SNE, Figure \ref{fig:tsne} reveals that most eviction judgements (marked by red pluses) and decisions (marked by green crosses) are concentrated in two distinct regions. 
To gain a deeper understanding of these two groups/clusters of eviction cases, we use LDA's output to uncover the most prevalent topics within each group. Figure \ref{fig:topic_hist_evict_clusters_all} provides a breakdown of topic prevalence within each group. The intensity of colour shading represents the proportion of cases within a cluster where a specific topic serves as the most prevalent. 
Figure \ref{fig:topic_hist_evict_clusters_all} shows that the larger eviction group (with 134 members) primarily consists of cases where `housing and property' dominates as the most prevalent topic for 70\% of cases (i.e.~94). 

In the smaller group (lower plot) with 36 cases, the primary topic for over 70\% of cases (i.e.~34) is `warfare'. After a closer look, we have observed that they are all related to the situation in southeastern Turkey during the 1990s where Turkish Armed Forces and paramilitary `village guards' systematically evacuated and destroyed rural settlements 
(see e.g., \textit{Nuri Kurt v.~Turkey} and \textit{Soylu v.~Turkey}).\footnote{ECHR 29 November 2005, no. 37038/97 (\textit{Nuri Kurt v.~Turkey}); ECHR 15 February 2007, no. 43854/98 (\textit{Soylu v.~Turkey}).} As such, the spatial separation of this smaller cluster from the other eviction cases is due to its specific association with cases related to this conflict, as opposed to the more prevalent and general eviction-related topic `housing and property' as found in the larger cluster.

In addition to these two groups, 28 eviction cases are scattered across various locations. These particular cases are all decisions except for two judgments. A comprehensive manual analysis, conducted by the authors with a legal background, has unveiled three factors that contribute to this dispersion of cases. The first factor is that some of these cases intertwine with another prominent legal issue other than eviction. The second factor relates to the focus on eviction within the context of Article 6 or other articles of the ECHR. The third factor is that some of these eviction cases are removed from the list of the ECtHR due to compensation or settlement agreements, and the text of the judgment/decision does not really deal with eviction in detail.

\subsection{Conclusions for Experiment 1}

In summary, through this experiment we have discovered that LDA can be a helpful tool for organizing and categorizing a large amount of case law. While topics identified by LDA may reveal previously unknown topics in ECtHR case law, the topics we identified (see Figure \ref{fig:topwords_all}) remain quite broad and correspond with topics discussed in the scientific literature on Article 8. LDA did not reveal more specific topics, such as evictions. Identifying these more specific subtopics would, however, be very useful for practitioners and researchers. We also observed that eviction case law was scattered in two larger topics, and LDA was not well-suited for identifying (i.e.~clustering) all cases on this specific subtopic. Consequently, if a researcher wants to focus on analyzing cases related to evictions, they may miss a small but interesting set of cases that are grouped with various other topics. Citation network analysis may offer a solution to this problem, and we will explore this approach in our next experiment.

\section{Experiment 2: Grouping cases within the citation network}\label{sec:net_analysis}
The citation network of court decisions could potentially reveal patterns in how cases connect and form groups of interconnected cases. These connected groups of cases are known as communities: clusters of nodes (i.e.~cases) that display robust interconnections, signifying a notable higher density of links within the group than connections outside of the group. Detecting and understanding these communities is a pivotal step in unravelling the structural complexities of a network. Experiment 2 therefore aims to identify communities of cases in the network of case law on Article 8. Thus, we apply a community detection algorithm to the giant subnetwork consisting of 7,234 cases (including both English and French cases; see Section \ref{subsec:cite_net}). In addition, we will locate the position of the cases of our manually annotated dataset on evictions and assess if these form multiple communities based on the type of eviction.

\subsection{Method}

One approach for identifying communities within a network involves comparing it to a random network \cite{barabasi2013network}.\footnote{In a random network, the chance of a connection between any pairs of nodes is equal This means there is no preference or underlying structure guiding the selection of neighbouring nodes for a given node.} The greater the dissimilarity between the network and a random model, the higher the likelihood that the network has an underlying structure.
The \textit{modularity} measure\cite{barabasi2013network} can be used to quantify the dissimilarity between the actual network and a random network by calculating and comparing the number of links within each community for both networks. 
To detect communities within the network, the modularity needs to be maximized. In determining the modularity, a choice needs to be made (via setting a so-called resolution parameter) between detecting either smaller or larger communities.

Various algorithms have been developed to maximize the modularity. Among these, the Louvain algorithm has been widely used \cite{blondel2008fast}. This is an iterative procedure in which first small communities are detected (by optimizing modularity) and successively larger communities are created by grouping these smaller communities. 

As the presence of a link in a citation network may represent the relatedness between different documents, we anticipate the existence of more connections among documents with similar contents. In the context of the citation network for Article 8 case law, this implies a direct relationship between the content of case law and its corresponding community. This highlights the importance of employing community detection methods. 

\subsection{Results: extraction of communities}
As we would like to group cases based on their citation behaviour, we focus on the large subnetwork consisting of 74\% (7,234) of the cases and 99.8\% (39,501) of the links between cases.\footnote{Note that other subnetworks only consist of 1 to 8 nodes (for more details see Section \ref{subsec:cite_net})}. 
To apply the Louvain algorithm, we need to set the resolution parameter. Here, we set it to two different values: the default value of 1 and a higher value of 3. 
Table \ref{tab:louvain_output} shows that the number of communities increases from 16 to 46 as we increase the resolution value. This shows the impact of the resolution on the resulting communities. A larger resolution value leads to more (and smaller) communities.

\begin{table}[]
    \centering
    \begin{tabular}{c|c|c c| c}
        \hline
        Resolution & Total & \multicolumn{2}{c|}{Communities} & Number of  \\
         & number of  & \multicolumn{2}{c|}{with eviction cases} & known evictions\\
         & communities& Name &Total community size& in community \\
        \hline
        \multirow{2}{*}{1}& \multirow{2}{*}{16} & (a)&494 & 105  \\
        & &(b) & 466& 46 \\
        \hline
        \multirow{2}{*}{3} & \multirow{2}{*}{46}& (a) & 218& 100  \\
        && (b1) & 264 & 24  \\
        && (b2) & 125 & 18  \\
        \hline
    \end{tabular}
    \caption{Details of communities within eviction cases}    \label{tab:louvain_output}
\end{table}

To gain a deeper understanding of each community's nature, a detailed assessment of all communities is necessary. Given the large number of cases (7,234), it is not feasible to perform such a comprehensive investigation. As a result, we narrowed our focus to a specific subset of 171 eviction-related cases within the giant subnetwork. The remaining 27 cases outside of this subnetwork are not considered in experiments 2 and 3.
Table \ref{tab:louvain_output} shows the details about the communities containing most of the known eviction-related cases. 

For a resolution of 1, the most eviction cases are concentrated in two communities. The first community (a) contains the majority of (i.e.~105) eviction cases. 
In the second community (b), we observe 46 eviction cases with 42 of them being related to the Kurdish-Turkish conflict. Additionally, there are 20 cases that are present in other communities.
These results are consistent with our findings from the topic modelling analysis conducted in Experiment 1: there seem to be two clusters, one related to general housing-related topics and another one related to the conflict in Turkey. 

By increasing the resolution parameter to 3, we observe the emergence of more (46) communities. Table \ref{tab:louvain_output} shows that most eviction cases are now distributed across three communities. Similar to the previous setting, the first community (a) contains the majority of (i.e.~100) eviction-related cases. 
However, the eviction cases about the Turkish conflict are now divided into two smaller communities. This division is based on document types as the third community (b2) only contains decisions, including 18 eviction-related cases. 
This result shows that using a higher resolution results in smaller communities. 

\subsection{Conclusions for Experiment 2}
In summary, similar to the result of the first experiment using LDA, we observe two eviction-related communities/clusters (using a resolution of 1). This is well-aligned with our expectation that the number of links among cases sharing a common topic is much higher than others (having different topics). A disadvantage of this approach is that the output of the Louvain algorithm strongly depends on the resolution value, which limits its application. In the next section, we show how the integration of textual information helps to overcome this challenge.

\section{Experiment 3: Integrating topic modelling and community detection} \label{sec:case_retrieval}

In the previous experiment, we saw how the dependency of the Louvain algorithm on the resolution value limited its ability to retrieve topic-focused communities.
In Experiment 3, we therefore assess whether a combination of the output of Experiment 1 (topic modelling) and the Louvain algorithm helps to reduce the impact of the resolution and to improve the ability of identifying communities that are more aligned with the issues represented within them. In this experiment, we therefore use the same citation network as in Experiment 2 with 7,234 (both English and French) cases.

\subsection{Method}

In Experiment 2, we assumed that all citations among cases have the same level of importance. In reality, however, the meaning of some citations from or to cases may be more relevant than others. Here, we show that the impact of resolution on the resulting communities can be reduced by letting links (or citations) have different levels of importance. 
In other words, we define scores (for each link) capturing the relevance of each citation. Later, we demonstrate how these scores help the Louvain algorithm in forming communities with similar themes. 

To measure the relevance of links, we utilize the contents of the cases. More precisely, we determine each link's relevance on the basis of the similarity of the topic vectors computed by LDA. In other words, the more similar the topics between two cases are, the stronger the relevance of the corresponding citation is. This weakens the strength of links between cases pertaining to different issues and strengthens the connection between cases with similar themes. As a result, this should help the Louvain algorithm identify more homogeneous communities. Specifically, the weight of a link is set equal to the cosine similarity between the topic vectors. Since we do not have topic vectors for French cases, we set the cosine similarity involving one or more French cases to the median cosine similarity (i.e.~0.755).

\begin{table}[]
    \centering
    \begin{tabular}{c|c|c c|c}
        \hline
        Resolution & Total& \multicolumn{2}{c|}{Community} & Number of \\
         &number of&\multicolumn{2}{c|}{with eviction}&  known evictions \\
        &communities& Name&Size & \\
        \hline
        \multirow{2}{*}{1} &\multirow{2}{*}{25}& (a)&248 &104 \\
        & &(b)&489 & 48  \\ 
        \hline
        \multirow{2}{*}{3}& \multirow{2}{*}{53} & (a)&209 & 102  \\
         & &(b)&209 & 38  \\ 
        \hline
    \end{tabular}
    \caption{Details of communities within eviction case using the weighted network. 
    }
    \label{tab:louvain_output_weighted}
\end{table}
This approach thus leads to a weighted citation network where a weight captures a link's relevance for the two cases. After assigning these weights, the Louvain algorithm can be applied to the (weighted) network to identify communities based on the strengths of connections.

\subsection{Results: eviction-related case retrieval}
Similar to Experiment 2, we focus on the large subnetwork containing 7,234 cases and 39,501 links. Moreover, we again set the resolution to two values: 1 and 3. Table \ref{tab:louvain_output_weighted}, shows that if we increase the resolution value (from 1 to 3), the number of communities increases from 25 to 53. If we compare these values to those results from Experiment 2, it is clear that incorporating the similarity scores results in the emergence of more communities (from 15 to 25 for a resolution of 1, and from 46 to 53 for a resolution of 3). This shows that the introduction of weights weakens the strength of some connections and, as a result, more communities are formed. 

Similar to Experiment 2, to conduct a more detailed analysis of community content, we focus on those communities containing the majority of eviction-related cases.
For the use of a resolution of 1, we observe a similar pattern as in the previous setting, with most eviction cases being concentrated in two communities. However, in the present experiment, we observe that the size of the first community (i.e.~248 cases) is half of the size of the corresponding community in Experiment 2 (i.e.~494 cases). The dominant topic for 47\% of cases within the community is 'housing and property', which is nearly twice the frequency of cases within the first community in the previous experiment\footnote{If we use the result of LDA to assign a primary topic to each case in the first community, obtained in Experiment 2, 'housing and property' is the primary topic for only 25\% of cases.}. 
This confirms the success of the introduced scores in weakening connections between cases with different topics. 

In Table \ref{tab:louvain_output_weighted}, we also observe that utilizing LDA helps to keep the unity of the second community when using a resolution of 3. This was not the case in Experiment 2, where this resolution resulted in two separate communities.  This result illustrates the benefit of our strategy in strengthening links between cases with similar themes.

\begin{table}[]
    \centering
    \begin{tabular}{c|c|c|c|c}
        \hline
        community & community & \# of English  & \# of housing & \# of eviction\\
        name&size&cases&related cases&related cases\\
        \hline
        (a) & 209&178& 151 & 113\\
        (b) & 209& 183& 131 & 98\\
        \hline
    \end{tabular}
    \caption{Detailed analysis of the communities in Table \ref{tab:louvain_output_weighted} for resolution of 3.}
    \label{tab:case_retrieval}
\end{table}

To assess the effectiveness of the derived communities in forming communities focusing on specific topics,
we conducted a manual, more in-depth analysis of all cases in English within both communities using a resolution of 3. Table \ref{tab:case_retrieval} reports the number of cases for both the manually assigned, more general topic of `housing' as well as the more specific topic `eviction'. In the first community (a) of connected cases, 85\% of (151 out of 178) English cases are found to be related to housing. Among them, 113 cases, including 23 newly detected ones (i.e.~not among the 198 manually identified cases), are found to deal with eviction. In the second community (b), we observe 131 (72\%) housing-related cases in the English language. In this community, we identified 98 cases as being directly associated with eviction, including 60 newly detected cases. This confirms the success of our approach to forming communities with more focus on a specific topic. 

Thus, this hybrid approach offers an effective way to retrieve cases focusing on a specific topic, such as eviction. For instance, here we can detect the first community, consisting of potential candidates related to eviction, since its primary topic (in 47\% of cases) is 'housing and property'. Furthermore, by tracking key cases such as \textit{Akdivar and others v.~Turkey},\footnote{ECHR 16 September 1996, no. 21893/93 (\textit{Akdivar and others v.~Turkey}).}, we can retrieve the second community. This leads to the total detection of 211 eviction-related cases by checking only 361 (=178+183) cases. Notably, this surpasses our initial manual query-based selection of 198 eviction cases. 
This result shows that our approach is useful to identify cases surrounding a similar topic.

\subsection{Conclusion for Experiment 3}
In summary, the results of Experiment 3 support the idea that incorporating LDA assists the Louvain algorithm in identifying more cohesive communities, as it can weaken the interconnections among cases addressing different issues and strengthen the association among cases addressing the same issues. This decreases the influence of the resolution value on the homogeneity of communities and provides an efficient way to use both textual and citation content for case law retrieval.

\section{Discussion and conclusion}\label{sec:conclusion}

In this closing section, we discuss the findings from our experiments designed to study and develop an enhanced computational method for finding, organizing and analyzing case law. We conducted three experiments using all case law on Article 8 of the ECHR from the European Court of Human Rights. In these three experiments, we applied: 1) topic modelling using LDA to cluster cases with similar topics, 2) citation network analysis using the Louvain Algorithm to form communities based on citation patterns, and 3) a combination of these two methods to uncover whether a combined approach leads to an enhanced method of finding and organizing cases. 
To scrutinize the level of detail revealed by these methods, we regularly zoom in on a subset of Article 8 cases that deal with one particular issue, namely eviction.

Our research reveals four key insights. First, our study highlights that while employing LDA for case law analysis yields a comprehensive overview of the overarching themes within the cases, it falls short in providing an exhaustive exploration of more specific subtopics, such as eviction. We found that Article 8 cases span seventeen distinct topics, encompassing diverse subjects from warfare to healthcare. Focusing on eviction cases, we found them primarily within the `housing and property rights' cluster, but a substantial portion remained scattered among other topics, or was embedded in the `warfare’ cluster, related to the conflict between the Turkish government and the Kurdish population (often resulting in the demolition of homes). For legal researchers seeking to identify eviction cases, the `housing and property rights' cluster appears to be the most logical choice. However, our study uncovers a potential pitfall in solely relying on LDA clustering, as it might inadvertently overlook a significant proportion of eviction cases.

Second, shifting our approach to network analysis using the Louvain algorithm did not solve the potential oversight of eviction cases. Our findings even resulted in identifying fewer eviction cases than with LDA. Due to the vast number of isolated nodes, we narrowed our focus to the giant subnetwork, housing 74\% of Article 8 cases, within which 87\% of our subset of eviction cases were found. As such, a portion of eviction cases falls outside the purview of our analysis. Using the Louvain algorithm and using different resolution parameters to analyze the giant subnetwork caused a further decrease in the number of identified eviction cases. We discovered two communities with eviction cases. The first community encompassed a total of 494 cases, of which 105 were eviction-related, while the second community contained 466 cases, with 46 of them being eviction cases. The majority of these 46 cases were associated with the conflict in Turkey. This finding underscores an important caveat: without prior knowledge of this particular, somewhat atypical cluster of eviction cases, this method could inadvertently lead to the oversight of nearly 47\% of the total 198 eviction cases.

Third, while employing topic modelling and network analysis in isolation may not yield promising results in the quest to find and organize case law, the combination of these methods proves to be highly effective. The merged approach revealed two smaller and tightly interconnected communities, both with a total of 209 cases. The first community housed 102 eviction cases, while the second contained 38. Identifying these smaller communities with relatively larger percentages of eviction cases underscores the utility of strengthening the connections between cases based on their thematic similarities.

Although a considerable portion of the manually identified eviction cases remains undiscovered, the emergence of these relatively small yet robust communities, characterized by strong connections among the cases, offers a valuable resource for identifying \textit{new} cases related to eviction. By fusing topic modelling with community detection, we successfully identified 83 previously undiscovered eviction cases. As such, combining topic modelling with citation network analysis leads to the best results in retrieving and organizing case law.

In summary, our study shows that while manual queries for relevant case law can be time-consuming and prone to human error, relying solely on computational methods carries its own limitations. Notably, it can lead to the oversight of relevant cases primarily centred around distinct topics and those that lack connections within the case network and thereby fall outside of the bounds of the giant subnetwork. However, this does not imply that the automation of case law collection is unattainable. Rather, as we have shown in this paper, it underscores the need for a guided, well-informed approach and may help to supplement manual queries. Such an approach combines the strengths of computational techniques with legal expertise, ensuring a more effective and comprehensive exploration of case law.


\bibliographystyle{unsrt}  
\bibliography{references}

\end{document}